\documentclass[mathleft
]{an}
\usepackage{graphicx}
\usepackage{times}
\begin{document}

\Pagespan{789}{}
\Yearpublication{2009}%
\Yearsubmission{2009}%
\Month{09}%
\Volume{999}%
\Issue{88}%

\title{Transport of cosmic rays in the nearby starburst galaxy
  NGC\,253\,\thanks{This work is based on observations with the Effelsberg
    100m and the VLA radio telescopes. The Effelsberg telescope is
    operated by the Max-Planck Institut f\"ur Radioastronomie (MPIfR). The VLA
    (Very Large Array) is operated by the NRAO (National Radio Astronomy
    Observatory).}}

\author{V. Heesen\inst{1,3}\fnmsep\thanks{Corresponding author:
  \email{v.heesen@herts.ac.uk}\newline}
\and  R. Beck\inst{2}
\and M. Krause\inst{2}
\and R.-J. Dettmar\inst{3}
}
\titlerunning{Transport of cosmic rays in NGC\,253}
\authorrunning{Volker Heesen et al.}
\institute{
Centre for Astrophysics Research, University of
  Hertfordshire, Hatfield AL10 9AB, UK
\and 
Max-Planck-Institut f\"ur
  Radioastronomie, Auf dem H\"ugel 69, 53121 Bonn, Germany
\and 
Astronomisches Institut der Ruhr-Universit\"at Bochum,
  Universit\"atsstr. 150, 44780 Bochum, Germany}

\received{September 2009}
\publonline{later}


\abstract{%
  Radio halos require the coexistence of extra-planar cosmic rays and magnetic
  fields.  Because cosmic rays are injected and accelerated by processes
  related to star formation in the disk, they have to be transported from the
  disk into the halo. A
  vertical large-scale magnetic field can significantly enhance this transport.\\
  We observed NGC\,253 using radio continuum polarimetry with the Effelsberg and VLA telescopes.\\
  The radio halo of NGC\,253 has a dumbbell shape with the smallest vertical
  extension near the center. With an estimate for the electron lifetime, we
  measured the cosmic-ray bulk speed as $300\pm30\,\rm km\,s^{-1}$ which is
  constant over the extent of the disk. This shows the presence of a ``disk
  wind'' in NGC\,253. We propose that the large-scale magnetic field is the
  superposition of a disk $(r,\phi)$ and halo $(r,z)$ component. The disk
  field is an inward-pointing spiral with even parity. The conical (even) halo
  field appears in projection as an X-shaped structure, as observed
  in other edge-on galaxies.\\
  Interaction by compression in the walls of the superbubbles may explain the
  observed alignment between the halo field and the lobes of hot H$\alpha$-
  and soft X-ray emitting gas. The disk wind is a good candidate for the
  transport of small-scale helical fields, required for efficient dynamo
  action, and as a source for the neutral hydrogen observed in the halo.}

\maketitle

\section{Introduction}
Gaseous halos are a common property of star-for\-ming ga\-laxies (Dettmar
1992). How galactic halos are formed is still under debate. One scenario is
that star formation in the disk drives an outflow by releasing hot gas and
cosmic rays into the interstellar medium. Their combined pressure can drive
the outflow against the gravitational pull of the stellar disk. Cosmic rays are particularly well suited because their adiabatic index of
$\Gamma=4/3$ results in a larger pressure scaleheight than that of the hot gas
($\Gamma=5/3$). Moreover, cosmic rays do not lose their energy fast by
radiation losses, so that they are able to contribute a significant pressure even
far away from the disk. Breitschwerdt at al.\ (1991) showed that the
cosmic-ray pressure is able to accelerate the flow in the halo, so that it can
escape from the gravitational potential as a galactic wind. A galactic wind,
hybridly driven by the cosmic-ray and thermal gas, may be present in the Milky
Way (Everett et al.~2008).
The structure of the large-scale magnetic field significantly influences the
cosmic-ray transport. A vertical field enhances the transport from the disk
into the halo, because the diffusion coefficient along the field lines is
larger than perpendicular to them. Vertical magnetic field components have
indeed been found in the halos of several nearby edge-on galaxies. They all
show an X-shaped field pattern like in NGC\,5775 (T\"ullmann et al.~2000) and
in NGC\,891 (Krause 2007). How these fields are formed is still under
debate. The geometry of the flow is shaped by the gravitational potential and
the pressure balance in the halo, so that it can have a radial structure
(Fichtner et al.~1991). The magnetic field lines may trace the direction of
the flow which has vertical and radial components, so that in projection it
forms an X-shape. Hydro simulations of disk galaxies with star formation show
also an X-shaped gas flow (Dalla Vecchia \& Schaye 2008).
Another question  is how large-scale magnetic fields are
generated in galaxies. The $\alpha$-$\Omega$ dynamo is among the favorite
explanations but it remains to be clarified how it wor\-ks in presence of a
galactic wind. One requirement is that small-scale helical fields are expelled
from the galaxy (Sur et al.~2007). Parker (1992) pointed out the importance of
cosmic rays to drive a fast dynamo where the magnetic field amplification is
enhanced by magnetic reconnection. Global MHD simulations of cosmic-ray driven
galactic dynamos show that field amplification is possible and that the
X-shaped structure can be reproduced (Hanasz et al.~2009a,b).
NGC\,253 is a nearby starburst galaxy which allows us to study its structure
in high detail ($D=3.94\,\rm Mpc$, $30\arcsec = 600\,\rm pc$). Is has one of
the brightest radio halos known, but so far a vertical magnetic field could
not be found (Beck et al.~1994). It has a prominent gaseous halo that is
visible in soft X-ray, H$\alpha$, and \ion{H}{I} emission (Bauer et al.~2008;
Hoopes et al.~1996; Boomsma et al.~2005) which makes it one of the best
candidates besides M\,82 for a superwind (Heckman et al.~2000).
\begin{figure}[t]
  \resizebox{1.0\hsize}{!}
{\includegraphics{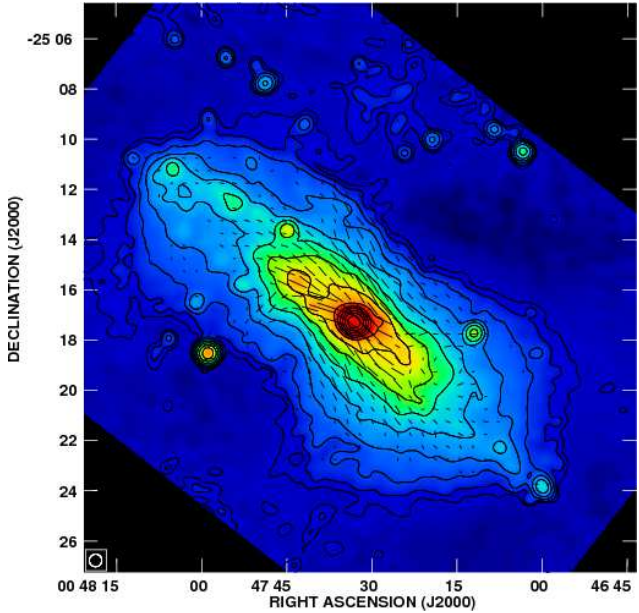}}
  \caption{Total power radio continuum at $\lambda 6.2\,{\rm cm}$
  obtained from the combined VLA + Effelsberg observations, smoothed to $30\arcsec$
  resolution. Contours are at 3, 6, 12, 24, 48, 96, 192, 384, 768,
  1536, 3077, 6144, 12288, and 24576 $\times$ the rms noise of $30\,{\rm\mu
    Jy/beam}$. The overlaid vectors indicate the orientation of the
  Faraday-corrected ordered magnetic field. A vector length of $1\arcsec$ is
  equivalent to $12.5\,{\mu\rm Jy/beam}$  polarized intensity.}
  \label{fig:vla6_tp}
\end{figure}

\section{Observations}
We observed NGC\,253 in the radio continuum polarimetry mode using the
Effelsberg 100m single-dish telescope and the VLA interferometer in
D-configuration. A deep VLA mosaic at $\lambda 6.2\,\rm cm$ was combined with
100-m Effelsberg observations in order to fill-in the missing zero-spacing
flux. Moreover, we observed with the Effelsberg telescope at $\lambda$ $3.6\,\rm
cm$ and used VLA maps from the archive at $\lambda\lambda$ $20\,\rm cm$ and
$90\,\rm cm$. Details of the observations and data reduction can be found in
Heesen (2008) and Heesen et al.~(2008; 2009).
\section{Cosmic-ray transport}
\begin{figure}[htb]
  \resizebox{1.0\hsize}{!}{\LARGE\input{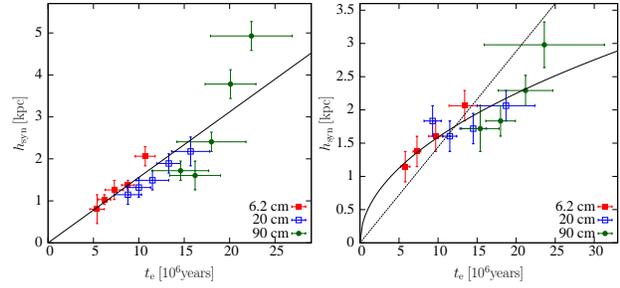}}
 \caption{Scaleheight $h_{\rm syn}$ of the thick radio disk as a
    function of the electron lifetime $t_{\rm e}$ in the northeastern
    halo (left) and in the southwestern halo (right). The linear fit is
    the theoretical expectation for a convective cosmic-ray transport
    with a constant bulk speed. The fit $h_{\rm syn}\propto\sqrt{t_{\rm e}}$ is the
    theoretical expectation for a diffusive cosmic-ray transport.}
  \label{fig:h_sl_col}
\end{figure}
Cosmic-ray electrons undergo several energy losses, among them loss due to
synchrotron radiation is dominant. The synchrotron energy loss depends on the
electron energy and the magnetic field strength as $dE/dt\propto E^2
B^2$. Electrons with higher energies have a shorter lifetimes, which causes a
steepening of the electron spectral index with time, known as electron aging
(e.g.\ Longair 2008). The electron spectral index $\gamma$ is related to the
radio spectral index by $\alpha=(\gamma-1)/2$. When electrons are injected and
accelerated in the disk, they have a flat spectral index of $\gamma\simeq
2$. In the halo, the electron population is aged and the spectral index is
steep. We now use the electron lifetime to measure the cosmic-ray bulk speed.
In Fig.\,\ref{fig:vla6_tp} we present the distribution of the total power
emission at $\lambda 6.2\,\rm cm$ at a resolution of $30\arcsec$ together with
the orientation of the ordered magnetic field.\footnote{All resolutions are
  referred to as the half power beam width (HPBW).} Note the dumb\-bell-shaped
halo that has its smallest vertical extent near the center of the galaxy. The
vertical profiles of the emission can be characterized by a two-component
exponential distribution, which is the superposition of a thin and a thick
radio disk. The thin radio disk corresponds to the star-forming disk and has a
scaleheight of $0.4\,\rm kpc$. The thick radio disk corresponds to the
electrons in the halo with a mean scaleheight of $1.7\,\rm kpc$. The
scaleheight is a useful measure to characterize the radio halo, because it is
almost independent of the sensitivity and resolution of the observations. We
found that the scaleheight depends on the galactocentric radius with a minimum
near the center whereas it is larger further out.
We determined the equipartition magnetic field stren\-gths using the revised
formula by Beck \& Krause (2005) with a proton to electron ratio of
$K=100$. The magnetic field strength in the disk is between $7\,\rm \mu G$ and
$18\,\rm \mu G$. The magnetic field strength is highest near the center of
the galaxy where the electron lifetime is smallest. We now define the average
cosmic-ray bulk speed as
\begin{equation}
  v = \frac{h_{\rm e}}{t_{\rm e}},
\end{equation}
where $h_{\rm e}$ is the electron scaleheight ($h_{\rm e}=2 h_{\rm syn}$,
assuming equipartition) and $t_{\rm e}$ is the electron lifetime. In
Fig.\,\ref{fig:h_sl_col} the synchrotron scaleheight is shown as a function of
the electron lifetime. In the northeastern halo we find a linear
dependence between the electron scaleheight and lifetime. This is indicative
of convective transport of cosmic rays and the magnetic fields. The bulk speed
is remarkably constant indicating the presence of a ``disk wind'' with an
average cosmic-ray bulk speed of $300\pm30\,\rm km\, s^{-1}$. In the
southeastern halo the electron scaleheight is proportional to
$\sqrt(t_{\rm e})$ which is indicative of diffusive cosmic-ray transport. The
diffusion coefficient is $\kappa=2.0\pm0.2\times 10^{29} \rm cm^2\, s^{-2}$.
\begin{figure*}
\resizebox{1.0\hsize}{!}{\includegraphics{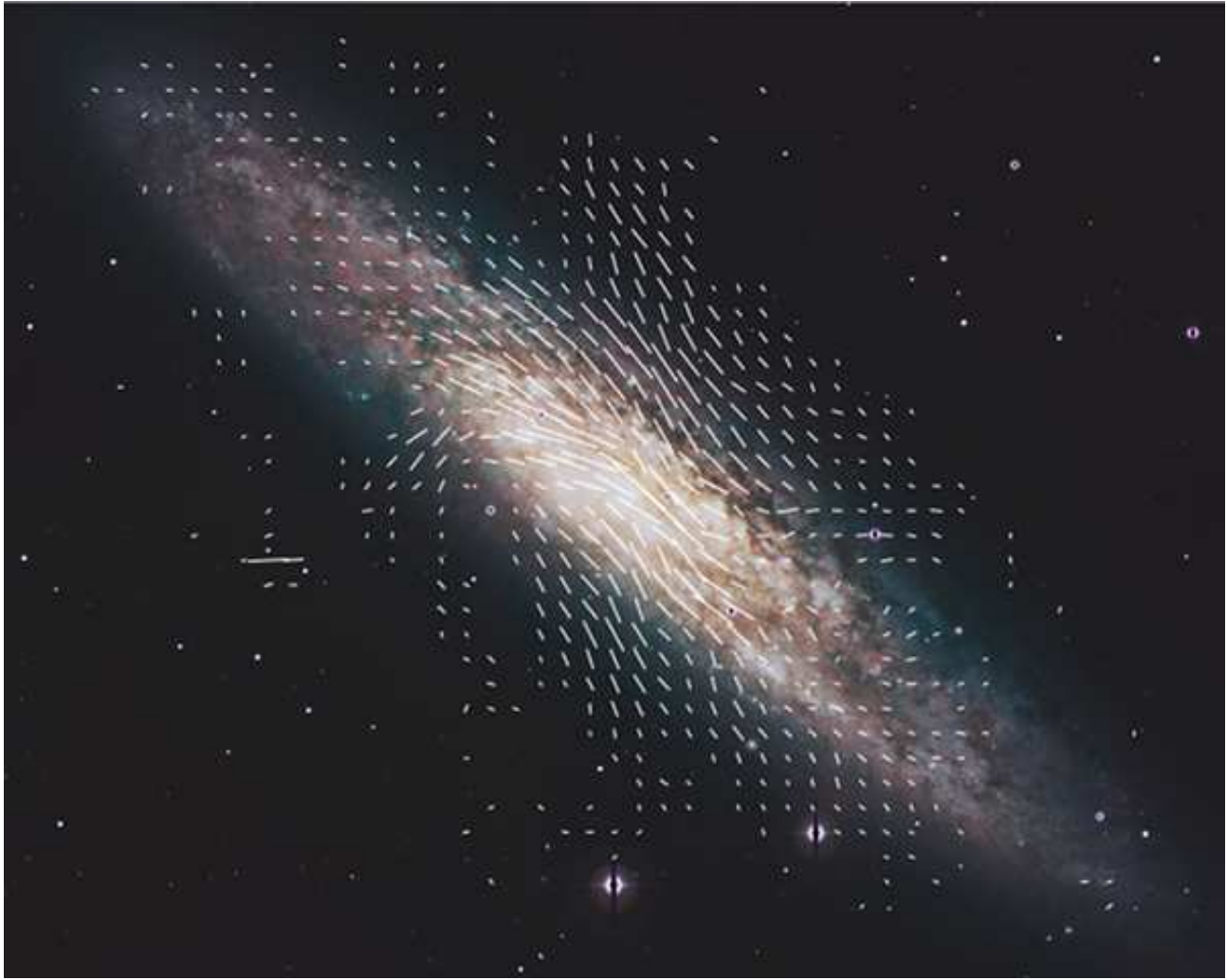}}
\caption{Orientation of the large-scale magnetic field at $\lambda 6.2\,\rm
  cm$ superimposed on an optical image by the ESO widefield imager. Image
  courtesy of NRAO/AUI.}
  \label{fig:nrao}
\end{figure*}
\section{Magnetic field structure}
\begin{figure}
  \resizebox{1.0\hsize}{!}{\includegraphics{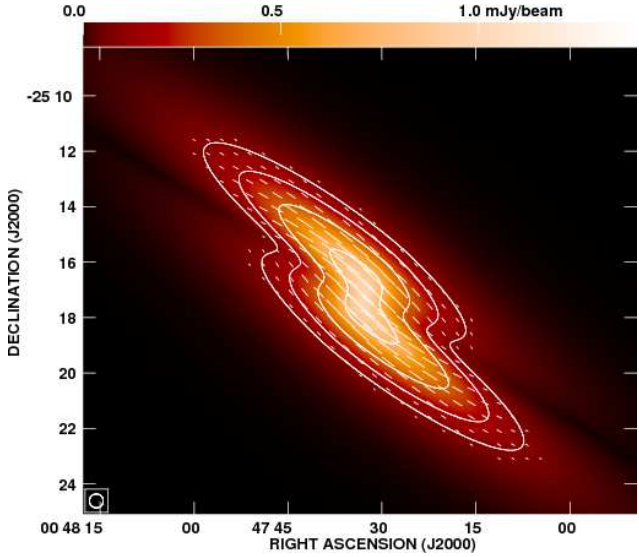}}
  \caption{Modeled polarized intensity at $\lambda 6.2\,\rm cm$ of the
    axisymmetric magnetic field (ASS) of the disk at $30\arcsec$
    resolution. Contours are at 3, 6, 12, and 24
    $\times$ $30\,{\mu\rm Jy/beam}$ of polarized intensity. The vectors
    indicate the magnetic field orientation where the length of the
    vectors of $1\arcsec$ is equivalent to $12.5\,{\mu\rm
      Jy/beam}$ polarized intensity.}
 \label{fig:disk}
\end{figure}

The high linear polarization of the synchrotron emission allows us to
measure the orientation of the large-scale magnetic
field. Figure \ref{fig:nrao} shows the magnetic field orientation in NGC\,253
superimposed on an optical image. The magnetic field vectors are mainly
parallel to the major axis but in some places we find a significant vertical
magnetic field component. These are at the locations of the so-called ``radio
spurs'', where the most prominent one is located east of the nucleus. We
propose that the large-scale magnetic field is the superposition of a disk
$(r,\phi)$ and a halo $(r,z)$ component.
In order to test this proposition we created a model for the disk magnetic
field using a spiral with a prescribed pitch angle of $25\degr$, similar to
the optical pitch angle of $20\degr$ (for details of the modeling see Heesen
et al.~2009). Spiral fields are observed in the disk of many face-on spiral
galaxies like M\,51 (Patrikeev et al.~2006). The model shown in
Fig.\,\ref{fig:disk} resembles the observed magnetic field well, so that we
can presume that the disk magnetic field is dominating. It is not symmetric
with respect to the minor axis as the pitch angle shifts the regions with the
strongest emission, so that we observe an S-shaped distribution of polarized
emission. The vertical field in the radio spurs is strong, so that the
emission of the disk is depolarized, otherwise the halo field is hidden by the
dominant disk field. This confirms our proposed two-component field structure.
\begin{figure}
  \resizebox{1.0\hsize}{!}{\includegraphics{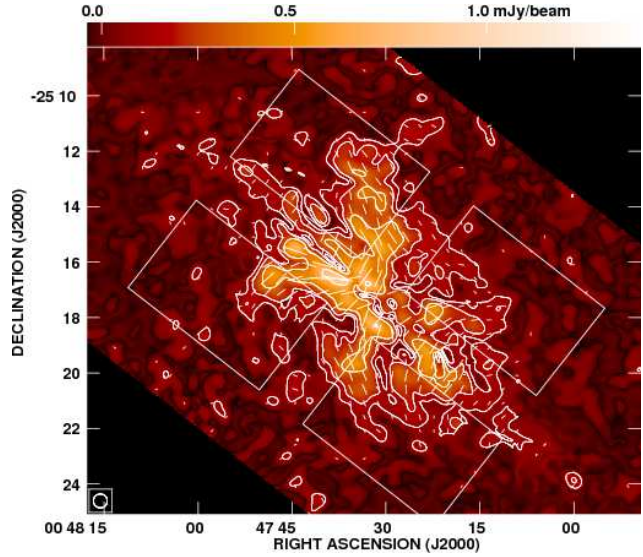}}
  \caption[]{Polarized intensity at $\lambda 6.2\,\rm cm$ of the halo
    magnetic field at $30\arcsec$ resolution after subtraction of the
    disk field model. Contours and vectors are identical to
    Fig.\,\ref{fig:disk}. The boxes for the integration of the
    orientation angle $\widehat\psi$ are also shown.}
  \label{fig:halo}
\end{figure}
If we subtract the disk magnetic field from the observations we retain the
halo magnetic field shown in Fig.\,\ref{fig:halo}. It has a prominent X-shaped
distribution in polarized intensity and a magnetic field orientation centered
on the nucleus. Such magnetic field structures have been observed in several
nearby edge-on galaxies (Dahlem et al.~1997; T\"ullman et al.~2000; Krause et
al.~2006; Krause 2007). But this is the first example for a galaxy which is
only mildly edge-on ($i=78.5\degr$) and is dominated by its disk magnetic
field.
The halo magnetic field has an average orientation angle of
$\widehat\psi=46\degr\pm15\degr$ with respect to the major axis. We propose a
conical field with a total opening angle of $\simeq 90\degr$ that appears as
an X-shaped field in projection.

\section{Magnetic field direction}
\begin{figure}
  \resizebox{1.0\hsize}{!}{\includegraphics{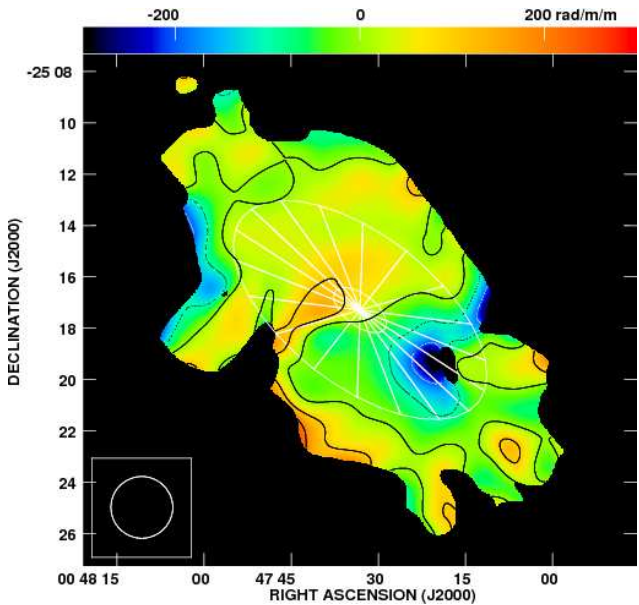}}
  \caption{RM distribution between $\lambda\lambda$ $6.2\,{\rm cm}$ and
    $3.6\,{\rm cm}$ from Effelsberg observations at $144\arcsec$
    resolution. Contours are at $-200$, $-100$, $0$, $100$, and $200$
    $\times$ $1\,{\rm rad\, m^{-2}}$. The vectors at both
    wavelengths were clipped below $4\times$ the rms noise level in
    polarized intensity prior to the combination.  The sectors for
    the azimuthal RM variation are also shown (Heesen et al.~2009).}
  \label{fig:eff_rm}
\end{figure}

\begin{figure}
  \resizebox{1.0\hsize}{!}{\includegraphics{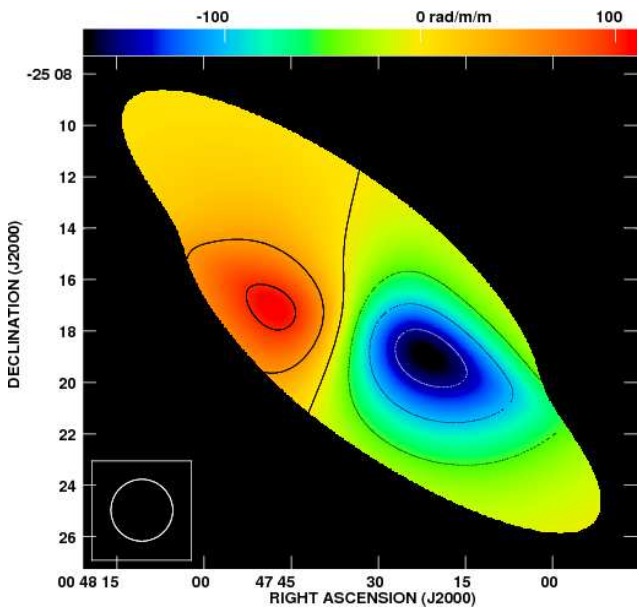}}
  \caption{RM distribution of the axisymmetric spiral (ASS) of the disk magnetic
    field at $144\arcsec$ resolution. Contours are at $-150$, $-100$, $-50$,
    $0$, $50$, and $100$ $\times$ $1\,\rm rad\,m^{-2}$. The map was clipped at
    $0.2\,\rm mJy/beam$ in polarized intensity.}
  \label{fig:diskrm}
\end{figure}

So far we have information about the orientation of the lar\-ge-scale
magnetic field in the plane of the sky only. Furthermore, Faraday rotation can tell us
the direction and the stren\-gth of the magnetic field along the
line-of-sight. We determine the Rotation Measure (RM) using the
difference of the polarization angles between two wavelengths:
\begin{equation}
  \Delta\chi={\rm RM}\cdot(\lambda_1^{\phantom{1}2}-\lambda_2^{\phantom{2}2}).
  \label{eq:mf_rm}
\end{equation}
In Fig.\,\ref{fig:eff_rm} we present the observed RM distribution calculated
from the Effelsberg observations at a resolution of $144\arcsec$. The
northeastern half contains positive RMs (magnetic field pointing to the
observer) and the southwestern half negative RMs (pointing away). The sense of
the radial rotation velocity is opposite to the RM which is the signature of
an inward-pointing magnetic field (Krause \& Beck 1998).
Based on the disk magnetic field model (Fig.\,\ref{fig:disk}) we created a
model RM distribution with a spiral magnetic field pointing inward. In case
of a Faraday screen where the polarized emission of a background source passes
the electrons and the magnetic field, the RM can be expressed by $RM=\int
n_{\rm e} B_{\parallel} ds$. The azimuthal RM variation of weakly inclined
galaxies can then be expressed as a cosine function with a phase shift
determined by the pitch angle (Krause et al.~1989). However, NGC\,253 is highly inclined (although
not edge-on), so that Faraday depolarization effects along the line-of-sight
can not be neglected (Sokoloff et al.~1998).  Therefore, we use a model to
calculate the RM along the line-of-sight which includes Faraday
depolarization. The m\-odeled RM distribution shown in Fig.\,\ref{fig:diskrm}
can reproduce the observed asymmetric RM distribution with different
amplitudes of the maximum and minimum.
According to our model, the direction of the disk magnetic field is equal on
both sides of the galactic midplane. Therefore, the disk field has even
parity. In order to study the direction of the halo magnetic field, we made a
model for the combined disk and halo field. We tested several possible
combinations of field directions in the disk and halo. An even halo field that
points away from the disk in the northern and the southern halo fits best to
the observations. However, we cannot safely exclude an odd halo field, because
we cannot reliably determine the field direction in the northern halo.

\section{Galactic wind}
\begin{figure}
  \resizebox{1.0\hsize}{!}{\includegraphics{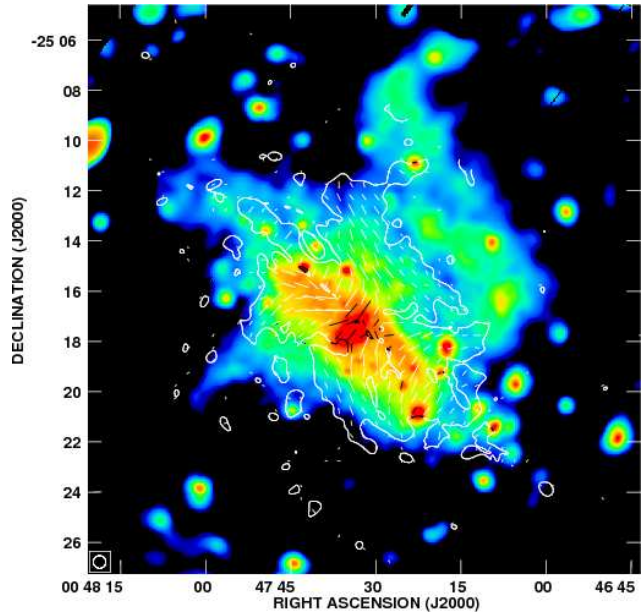}}
  \caption{Halo magnetic field overlaid onto diffuse X-ray emission. The contour
    is at 3 $\times$ $30\,{\mu\rm Jy/beam}$ (the rms noise
    level). A vector length of $1\arcsec$ is equivalent to $12.5\,{\mu\rm
      Jy/beam}$ polarized intensity. The X-ray map is from XMM-Newton
    observations in the energy band $0.5-1.0\,\rm keV$ (Bauer et al. 2008).}
  \label{fig:halo_xmm2}
\end{figure}

\begin{figure}
  \resizebox{1.0\hsize}{!}{\includegraphics{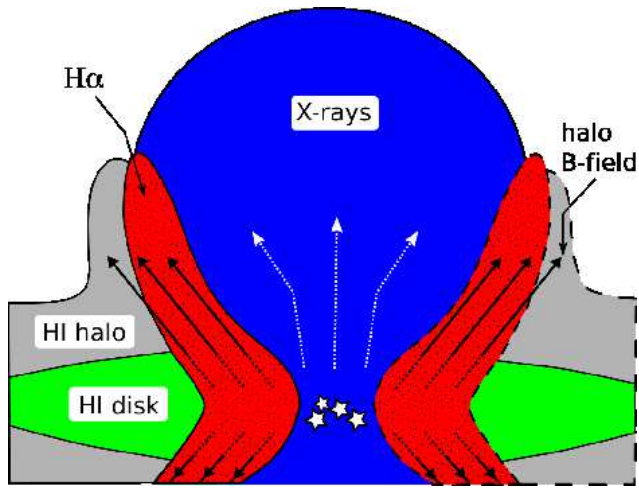}}
  \caption{Halo structure of NGC\,253. Reproduced from
    Boomsma et al. (2005) and extended. The superbubble, filled with soft
    X-rays emitting gas, expands into the surrounding medium (indicated
    by dotted lines with arrows). The halo magnetic field is aligned
    with the walls of the superbubble. Dashed lines denote components
    that are not (or only weakly) detected in the southwestern half of
    NGC\,253.}
  \label{fig:superbubble_even}
\end{figure}

In the halo of NGC\,253 we find emission from various spe\-cies of the
interstellar medium. Plumes of H$\alpha$, soft X-ray, and \ion{H}{I} emission
are extending far into the halo. They are forming shell-like structures with
the soft X-ray emission innermost, the \ion{H}{I} emission outermost and the
H$\alpha$ in between (Strickland et al.~2002).  A fast wind of hot, low-density
gas from the nuclear starburst can transport the material into the halo. Such
winds are observed in several starburst galaxies and are known as superwinds
(Heckman et al.~2000). A second possibility for the transport is the disk wind
that is traced by the cosmic rays over the full extent of the disk.
How does the magnetic field relate to the halo structure? In
Fig.\,\ref{fig:halo_xmm2} we show the halo magnetic field as an overlay of
soft X-ray emission. The magnetic field is concentrated near the center of
the galaxy. Its outer boundary along the major axis is near to the lobes seen
in the X-ray emission. Moreover, the orientation of the halo magnetic field is
tangential to the lobes. We note that the continuum emission
(Fig.\,\ref{fig:vla6_tp}) does not show any connection to the hot gas but only
the polarized emission. The interaction with the superbubble may align and
amplify the magnetic field in the shells of the superbubble. Moreover, the
magnetic field also shows geometrical limb brightening, because the
line-of-sight is perpendicular to the magnetic field lines in the shells. The
halo structure together with the magnetic field is shown in
Fig.\,\ref{fig:superbubble_even}.
The field lines in the halo may wind up in a spiral caused by the differential
rotation. This field structure is similar to the Parker spiral of the solar
magnetic field. On the other hand, if the transport of gas, field, and angular
momentum is large, the field lines are expected to corotate far into the halo
with no azimuthal component, as observed. The galactic wind can transport
angular momentum far better than the superwind in the center at small galactic
radii (Zirakashvili et al.~1996). The disk wind is also able to transport the
small-scale helical magnetic fields, as requested for efficient dynamo
amplification (Sur et al.~2007).
\section{Conclusions}
NGC\,253 has a large radio halo. The analysis of the scaleheights of the
cosmic-ray electrons and their lifetime shows that this galaxy possesses a
disk wind with a cosmic-ray bulk speed of about $300\,\rm km\,s^{-1}$. The
magnetic field consists of a disk and a halo component. The halo component
becomes visible after subtracting the disk and shows a prominent X-shape as
known from edge-on galaxies. This is the first X-shaped field in a galaxy
which is only mildly edge-on. The disk wind can effectively transport gas from
the disk into the halo. The observed magnetic field structure can be explained
by the interaction with the superwind by compression in the walls of the
superbubble and by limb brightening. Furthermore, the disk wind can transport
the small-scale helical fields to ensure efficient dynamo action.
\acknowledgements The organizers of the Splinter Meeting are thanked for
organizing an enjoyable and fruitful meeting.\\
VH acknowledges the funding by the Graduiertenkolleg GRK\,787 and the
Sonderforschungsbereich SFB\,591 during the course of his PhD.  The GRK\,787
``Galaxy groups as laboratories for baryonic and dark matter'' and the
SFB\,591 ``Universal properties of non-equilibrium plasmas'' are funded by the
Deutsche Forschungsgemeinschaft (DFG). RJD is supported by DFG in the
framework of the research unit FOR\,1048.\\
We thank Michael Bauer for providing the XMM-Newton map and Birgitta Burggraf
and Olaf Schmith\"usen are thanked for providing the map from the ESO
widefield imager.


\begin{thebibliography}{}
\bibitem{} Bauer, M., Pietsch, W., Trinchieri, G., Breitschwerdt, D., Ehle,
  M., Freyberg, M.~J., Read, A.~M.: 2008, A\&A 489, 1029
\bibitem{} Beck, R., Carilli, C.~L., Holdaway, M.~A., \& Klein, U.: 1994, A\&A
  292, 409
\bibitem{} Beck, R. \& Krause, M.: 2005, AN 326, 414
\bibitem{} Breitschwerdt, D., McKenzie, J.~F., \& V\"olk, H.~J.: 1991, A\&A
  245, 79
\bibitem{} Breitschwerdt, D., Dogiel, V.~A., V\"olk, H. J..: 2002, A\&A 385,
    216
\bibitem{} Boomsma, R., Oosterloo, T., Fraternali, F., van der Hulst J.~M.,
    Sancisi, R.: 2005, A\&A 431, 65
\bibitem{} Dahlem, M., Petr, M.~G., Lehnert, M.~D. et al.: 1997, A\&A 320, 731
\bibitem{} Dalla Vecchia, C. \& Schaye, J.: 2008, MNRAS 387, 1431
\bibitem{} Dettmar, R.-J.: 1992, Fundamentals of Cosmic Physics 15, 143
\bibitem{} Everett, J.~E., Zweibel, E.~G., Benjamin, R.~A. et al.: 2008, ApJ 674, 258
\bibitem{} Fichtner, H., Neutsch, W., Fahr, H.~J., Schlickeiser, R.: 1991,
  ApJ 371, 98
\bibitem{} Hanasz, M., Otmianowska-Mazur, K., Kowal, G., Lesch, H.: 2009a,
  A\&A 498, 335
\bibitem {} Hanasz, M., W\'olta\'nski, D., Kowalik, K.: 2009b, submitted, e-print arXiv:0907.4891
\bibitem{} Heckman, T.~M., Lehnert, M.~D., Strickland, D.~K., \& Armus, L.: 2000,
  ApJS 129, 493
\bibitem{} Heesen, V.: 2008, PhD thesis, Ruhr-Universit\"at Bochum, Germany
\bibitem{} Heesen, V., Beck, R., Krause, M., \& Dettmar, R.-J.: 2008, A\&A 494,
  563
\bibitem{} Heesen, V., Krause, M., Beck, R., \& Dettmar, R.-J.: 2009, A\&A
  accepted, e-print arXiv:0908.2985
\bibitem{} Hoopes, C.~G., Walterbos, R.~A.~M., \& Greenwalt, B.~E.: 1996, AJ 112,
  1429
\bibitem{} Krause, F. \& Beck, R.: 1998, A\&A 335, 789
\bibitem{} Krause, M.: 2007, Memorie della Societa Astronomica Italiana, Vol.\
  78, 314 
\bibitem{} Krause, M.: 2008,  in Magnetic Fields in the Universe II, ed.\ A.\
  Esquivel, Rev.\ Mex.\ Astron.\ Astrof.\ (SC), arXiv:astro-ph/0806.2060v2
\bibitem{} Krause, M., Hummel, E., \& Beck, R.: 1989, A\&A 217, 4
\bibitem{} Krause, M., Wielebinski, R., \& Dumke, M.: 2006, A\&A 448, 133
\bibitem{} Longair, M.: 2008, High Energy Astrophysics, Volume 2, 2nd edition,
  Cambridge University Press
\bibitem{} Patrikeev, I., Fletcher, A., Stepanov, R. et al.: 2006, A\&A 458, 441
\bibitem{} Sokoloff, D.~D., Bykov, A.~A., \& Shukurov, A. et al.: 1998, MNRAS
  299, 189
\bibitem{} Strickland, D.~K., Heckman, T.~M., \& Weaver, K.~A. et al.: 2002, ApJ 568, 689
\bibitem{} Sur, S., Shukurov, A. \& Subramaniam, K.: 2007, MNRAS 377, 874
\bibitem{} T\"ullmann, R., Dettmar, R.-J., \& Soida et al.: 2000, A\&A 364, L36
\bibitem{} Zirakashvili, V.~N., Breitschwerdt, D., Ptuskin, V.~S., V\"olk,
  H.~J.: 1996, A\&A 311, 113
\end{thebibliography}
\end{document}